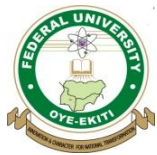

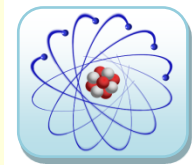

# Estimating All-Net Radiation using routine Meteorological data in Akure, Nigeria

## A.H. Olabintan[1*], E. Lawan[1] and T.A Oluwadare[2]

[1]Department of Physics, Federal University Oye-Ekiti, Ekiti State, Nigeria
[2]Department of Chemistry, Federal University Oye-Ekiti, Ekiti State, Nigeria

**Corresponding author**:

adewale.olabintan@fuoye.edu.ng

**ORCID ID:** 0009-0004-9119-7650

**Phone No:** +2348036550941

**DOI:** 10.55518/fjpas.NIPX7389

## Abstract

Earth's net radiation, sometimes referred to as net flux, is the balance between incoming and outgoing solar radiation energy from the atmosphere. This study utilised five years of data on tropospheric variables (solar radiation, air temperature, surface temperature, maximum temperature, minimum temperature, and dew point temperature), from January 2007 to December 2011 to estimate net radiation and examine its temporal (diurnal) and daily variation. The measurement site is the old site of the Nigerian Television Authority, NTA Akure at Iju (7.15°N, 5.12°E), in Akure North Local Government area of Ondo State. The results indicate that the net radiation exhibits a radiant energy surplus during the daytime and a deficit at nighttime. The daily mean maximum of the hourly net radiation flux was about 550 Wm$^{-2}$ for the dry season. The value occurred around 12:00 hrs and 380 Wm$^{-2}$ for the wet season, the value occurring around 12:30 hrs. We observed a time lag between net radiation and temperature; the net radiative flux reached its maximum at 12:00 hrs while the temperature attained its peak value around 15:00 hrs. Daily averages of net radiation exhibited dual peaks with values of about 160 Wm$^{-2}$day$^{-1}$ around February and November and a minimum of about 50 Wm$^{-2}$day$^{-1}$. The daily average temperature has a singular peak in February, the hottest month. We also analysed daily net radiation values from 10 models that differ in structure and complexity, and a lot of them underestimated the standard model because they were location-dependent.

## 1.0    Introduction

The surface heat flux is an important input parameter in boundary layer and air pollution dispersion models. For the practical application of these models, simple estimates of the heat flux from routine synoptic data are useful. Such estimates, in principle, can be obtained from surface energy budget considerations. The first step in this approach is to estimate the net radiation, this can be done through measured incoming shortwave radiation, screen temperature and cloud cover [1], [2].

Cloud cover and the structure and composition of the atmosphere are major determinants of net radiation, a key characteristic that controls the climate of the lowest layers of the atmosphere. When evaluating the evolution of the climate from both local and global viewpoints, net radiation is essential in addition to surface characteristics like albedo, emissivity, temperature, humidity, and thermal properties of the underlying soil [3]. It is the primary energy source for biological and physical processes, including evapotranspiration, which is utilised to maximise crop quality and production, as well as the planning of water resources and weather forecasting [4], [5], [6], [7].

In both the tropical and subtropical regions, the absorbed solar radiation generally exceeds the outgoing long-wave radiation, resulting in a net radiative heating of the planet, whereas in the middle to polar latitudes, there is a net radiative cooling [8]. This equator-to-pole difference, or gradient in radiative heating, is the primary mechanism that drives the atmospheric and oceanic circulations [9]. Thus, the net radiation budget at the surface is affected by several factors, including geographic, atmospheric, and seasonal (temporal) features. The annual variation in net radiative energy follows the solar declination [10].

Net radiation can be represented by the algebraic sum of both the downward and upward components of shortwave and long-wave radiation flux densities:

$$R_n = (1-\propto)R_s + R_l \downarrow -R_l \uparrow \qquad (1)$$

Where $R_s$ is the total downward solar radiation flux density, $\propto$ is the surface albedo, $R_l \downarrow$ is the downward long-wave radiation flux and $R_l \uparrow$ is the upward long-wave radiation flux density. All fluxes are in $Wm^{-2}$, representing radiation through a horizontal plane [11], [12].

In tropical West Africa, net radiation data are scarce, as only a few data available for the region were collected from field experiments such as the HAPEX-Sahel [13] and other independent studies [14], [15], [16]. In situations where net radiation data are unavailable for air pollution dispersion models, obtaining estimates from the terms that appear in the net radiation equation is common practice [17], [18].

Studies at Osu, a site about 20 km from Ile-Ife, Nigeria, have shown that maximum net radiation flux occurs around 1400 hours local time, about 380 $Wm^{-2}$ for April-October and 480 $Wm^{-2}$ for November-March. The large seasonal variability is attributed to the crucial roles played by convective clouds and water vapour in the atmospheric radiation budget [16], [19], [20].

This work aims to estimate net radiation from an ongoing in-situ measurement of some tropospheric parameters through the ongoing experiment by the Communication/Atmospheric Physics Research Group of the Department of Physics FUTA Southwestern Nigeria. The results of the estimated net radiation obtained in this study will assist in characterising the land-atmosphere energy exchange processes in the surface layer over the region for the study of its micro-meteorology [21], [22], [23].

## 2.0 Methods and Materials

### 2.1 Data Collection

We used five years of tropospheric variables (solar radiation, air temperature, surface temperature, maximum temperature, minimum temperature, and dew point temperature) spanning from January 2007 to December 2011. The measurement site is located at the old site of the Nigerian Television Authority, NTA Akure at Iju, in Akure North Local Government, a few kilometres from the city of Akure.

The sensors are mounted close to the ground surface and at different heights on a communication mast. The receiver (Console), together with the data logger, is located in a measurement room adjacent to the location of the radiation sensors. The device used for the measurement is the Davis 6162 wireless Vantage Pro Plus, manufactured by Davis Instruments, Hayward, California, United States of America.

### 2.2 Estimation of Net Radiation

We used the parameterised relationship of Jegede *et al.* [24] to determine net radiation in this study. The net radiation ($R_n$) at the surface is given as the algebraic sum of the downward flux of emitted solar radiation ($R_s \downarrow$) from the sun and sky (global radiation), the downward infrared or thermal radiation flux ($R_i \downarrow$) from the atmosphere, the upward flux of reflected solar radiation ($R_s \uparrow$), and the upward infrared radiation flux ($R_i \uparrow$) from the surface.

The equation used to calculate net radiation for this study is:

$$R_{cloud} = \begin{cases} R_{clear}\beta & \text{for daytime} \\ 0.1\,R_{clear} & \text{for nighttime} \end{cases} \qquad (2)$$

Where $R_{clear}$ is the net radiation for clear skies and $\beta$ is the cloud amount ratio determined using measured $\left(R_{sd(meas)}\right)$ and calculated $\left(R_{sd(clear)}\right)$ clear sky global radiation.

### 2.3 Net Radiation Calculation Procedure for the 9 Other Net Radiation Models

This section provides brief background information and describes the general procedures and common equations used in the different $R_n$ methods for the calculation of various parameters and variables.

Model 1: ASCE-EWRI [25]

$$R_n = R_{ns} \downarrow - R_{nl} \uparrow \qquad (3)$$

where;

- $R_n$ = net radiation ($MJ\ m^{-2}\ d^{-1}$)

- $R_{ns}$ = incoming net shortwave radiation ($MJ\ m^{-2}\ d^{-1}$)

- $R_{nl}$ = outgoing net long-wave radiation ($MJ\ m^{-2}\ d^{-1}$).

The $R_{ns}$ is a result of the balance between incoming and reflected solar radiation as a function of:

$$R_{ns} = (1-\propto)R_s \downarrow \qquad (4)$$

where;

- $\propto$ = albedo or canopy reflection coefficient (fixed at 0.23 for a green vegetation surface)

- $R_s$ = total incoming shortwave solar radiation $\quad (MJ\ m^{-2}\ d^{-1})$

The rate of $R_{nl}$ is proportional to the fourth power of the absolute temperature of the surface:

$$R_{nl} = \sigma \left[\frac{T^4{}_{max,k} + T^4{}_{min,k}}{2}\right]\left(0.34 - 0.14\sqrt{e_a}\right)\left(1.35\frac{R_s}{R_{so}} - 0.35\right) \qquad (5)$$

where;

- $\sigma$ = Stefan-Boltzmann constant $(4.903 \times 10^{-9}\ MJ\ m^{-2}\ d^{-1})$

- $T^4{}_{max,k}$ = daily maximum absolute air temperature (K=°C+273.16)

- $T^4{}_{min,k}$ = daily minimum absolute air temperature (K=°C+273.16)

- $e_a$ = actual vapour pressure of the air (kPa)

- $R_{so}$ = Calculated clear-sky solar radiation $(MJ\ m^{-2}\ d^{-1})$.

The actual vapour pressure is calculated as:

$$e_a = 0.6108\ exp\left[\frac{17.27 T_{dew}}{T_{dew} + 237.3}\right] \qquad (6)$$

where $T_{dew}$ is the dew point temperature (°C). The clear-sky solar radiation in equation (5) is estimated as;

$$R_{so} = (0.75 + 2 \times 10^{-5}z)R_a \downarrow \qquad (7)$$

Daily extra-terrestrial radiation $R_a$ $(MJ\ m^{-2}\ d^{-1})$ can be calculated as a function of the day of the year, solar constant, declination and latitude:

$$R_a = \frac{1440}{\pi}G_{sc}\delta_r \times [\omega_s \sin(\varphi)\sin(\delta) + \cos(\varphi)\cos(\delta)\sin(\omega_s)] \qquad (8)$$

where;

- $G_{sc}$ = solar constant (0.0820 $MJ\ m^{-2}\ d^{-1}$)

- $\delta_r$ = inverse relative distance from the earth to the sun

- $\omega_s$ = sunset hour angle (rad)

- $\varphi$ = latitude (rad)

- $\delta$ = solar declination (rad)

$$\delta = 0.4093\ \sin\left[\frac{2\pi(284+J)}{365}\right] \qquad (9)$$

where J is the day of the year (1 to 366)

$$\delta_r = 1 + 0.033\cos\left(\frac{2\pi J}{365}\right) \qquad (10)$$

$$\omega_s = arc\cos(\psi) \qquad (11)$$

$$\psi = \tan(\varphi)\tan(\delta) \qquad (12)$$

$R_{nl}$ is majorly influenced by temperature and humidity, but besides these, it is also influenced by cloudiness and the difference between the surface's and the air's temperatures. $R_{nl}$ is calculated by, [26];

$$R_{nl} = f\varepsilon\sigma T^4 \qquad (13)$$

where $\sigma$ is the Stefan-Boltzman constant $(4.895 \times 10^{-9}\ MJ\ m^{-2}\ d^{-1}\ K^{-4})$, T is the average air temperature, f is a factor to adjust for cloud cover, and $\varepsilon$ is the atmospheric emissivity. Wright and Jensen *et al.*, [27] proposed the following equation for f:

$$f = a \frac{R_s}{R_{so}} + b \tag{14}$$

and Brunt [28] developed an empirical equation for atmospheric emissivity ε:

$$\varepsilon = a_1 + b_1 \sqrt{e_a} \tag{15}$$

where $a = 1.35$, $b = -0.34$, $a_1 = 0.35$, and $b_1 = -0.14$; $e_a$ is calculated using equation (6).

Wright [29] presented an approach for dynamic values of a, b, and $a_1$:

for      $R_s/R_{so} > 0.7$: $a = 1.126 \ and \ b = -0.07$

for      $R_s/R_{so} \leq 0.7$: $a = 1.017 \ and \ b = -0.06$

$$a_1 = 0.26 + 0.1 \exp \left\{ -\left(0.0154(30m + N - 207)\right)^2 \right\} \tag{16}$$

where m is the month, and N is the day of the year

Another equation, which requires only mean air temperature to estimate ε, was presented by [30]:

$$\varepsilon = -0.02 + 0.261 \exp[-7.77 \times 10^{-4}(273 - T)^2] \tag{17}$$

To compute f to adjust for cloud cover, one can obtain estimated values of Rso by plotting observed Rs values to obtain an envelope curve through the maximum radiation values. Allen *et al.* [31] presented two procedures to calculate Rso. If the calibrated values of $a_s$ and $b_s$ are available, equation (18) is recommended:

$$R_{so} = (a_s + b_s)R_a \tag{18}$$

Doorenbos and Pruitt [32] recommended using 0.25 and 0.50 for $a_s$ and $b_s$, respectively. Jensen *et al.* [26] developed a linear equation correlating $R_n$ with $R_s$ and is expanded as:

$$R_n = a_3 R_s + b_3 \tag{19}$$

where the regression coefficients, $a_3 = 0.61$ and $b_3 = -1.0$, were obtained by averaging data from 14 locations worldwide. Irmak *et al.*, [2] derived an equation to predict daily $R_n$ using minimum climatological data, only the first equation will be used, which requires $T_{max}$, $T_{min}$, and measured $R_s$, and $d_r$ as input parameters:

$$R_n = (-0.054T_{max}) + (0.111T_{min}) + (0.462R_s) + (-49.243d_r) + 50.831 \tag{20}$$

where the units of $T_{max}$, $T_{min}$, and measured $R_s$ are $^0$C, $^0$C, and $MJ \ m^{-2} \ d^{-1}$, respectively.

All models except models 8 and 9 use equation (3) to calculate $R_n$ as a difference between $R_{ns}$ and $R_{nl}$. $R_{ns}$ was calculated from the observed $R_s$ using equation (4), whereas equation (13) was used to estimate $R_{nl}$. Models 8 and 9 estimated $R_n$ directly from equations (19) and (20) respectively. Table 1 shows the summary of the structure of the models.

**Table 1**: Structure of the models used to compute $R_n$ as a function of variables, constants, and equations

| | | | | $R_{nl}$ (eq. 5) | |
| | $R_{ns}$ (eq. 3) | f (from eq 13) | | | $\varepsilon$ |
| Models | $R_s$ | $R_{so}$ | a and b | | $a_1$ and $b_1$ |
| 2 | Measured | Eq. (17) | $a = 1.35; b = 0.34$ | Eq. (16) | ---- |
| 3 | Measured | Eq. (17) | Variable | Eq. (14) | $a_1 = 0.35; b_1 = -0.14$ |
| 4 | Measured | Eq. (6) | Variable | Eq. (16) | --- |
| 5 | Measured | Eq. (6) | Variable | Eq. (14) | $a_1 = 0.35; b_1 = -0.14$ |
| 6 | Measured | -- | --- | Eq. (14) | $a_1 = 0.35; b_1 = -0.14$ |
| 7 | Measured | -- | --- | Eq. (16) | --- |

Note: Models 8 and 9 use equations (19) and (20) respectively to estimate $R_n$ directly so they are not included in the table

## 3.0 Results and Discussion

### 3.1 Diurnal and Seasonal Variations of Net Radiation and Solar Radiation

Figure 1 shows the diurnal variation of the net radiative flux during the peak of the two seasons (wet in July and dry in January); they show similar patterns, with lower daytime net radiation values in the wet seasons compared to dry seasons. The daily mean maximum of the hourly net radiation flux was about 550 $Wm^{-2}$ for the dry season (occurring around 12:00 hrs) and 380 $Wm^{-2}$ for the wet season (occurring around 12:30 hrs).

The nighttime values did not reflect remarkable differences between seasons but were negative for both, indicating that outgoing net long-wave radiation exceeded net incoming shortwave radiation during these hours. The significant difference in daytime net radiation between the two seasons suggests that the convective clouds predominant during the wet season in the tropics are more effective in altering the radiation balance than the turbid air of the dry season. These two effects can be regarded as almost separate manifestations because the thunderstorms of the wet season continuously wash out particulates. In contrast, the surface conditions during the dry season significantly inhibit cloud formation.

Similar trends were observed in solar radiation flux, as shown in Figure 2, with daily mean maximum values of about 610 $Wm^{-2}$ for the dry season (occurring around 12:00 hrs) and 420 $Wm^{-2}$ for the wet season (occurring around 12:30 hrs). This similarity in patterns between net radiation and solar radiation flux indicates that solar radiation flux largely determines the net radiation flux in the tropics.

We observed a time lag between net radiation and temperature, as shown in Figure 3 for the dry season and Figure 4 for the wet season,

with net radiative flux peaking at 12:00 hrs while temperature peaked around 15:00 hrs. This occurs primarily because of the delay in heating of the atmospheric column by incoming solar radiation [10]. When the surface becomes heated, the adjoining atmosphere gains heat by convection, leading to an increase in air temperature. The dry season recorded higher temperature values than the wet seasons, with peak values reaching 31°C for the dry season and 27°C for the rainy season.

Relative humidity showed an inverse relationship with temperature, peaking during the early morning hours and recording its lowest values around 16:00 hrs, as illustrated in Figures 5 and 6. As the temperature increased after sunrise, relative humidity decreased, reaching its lowest point when the maximum temperature was realised. Relative humidity for the wet season was higher than that for the dry season, as can be observed by comparing Figures 5 and 6.

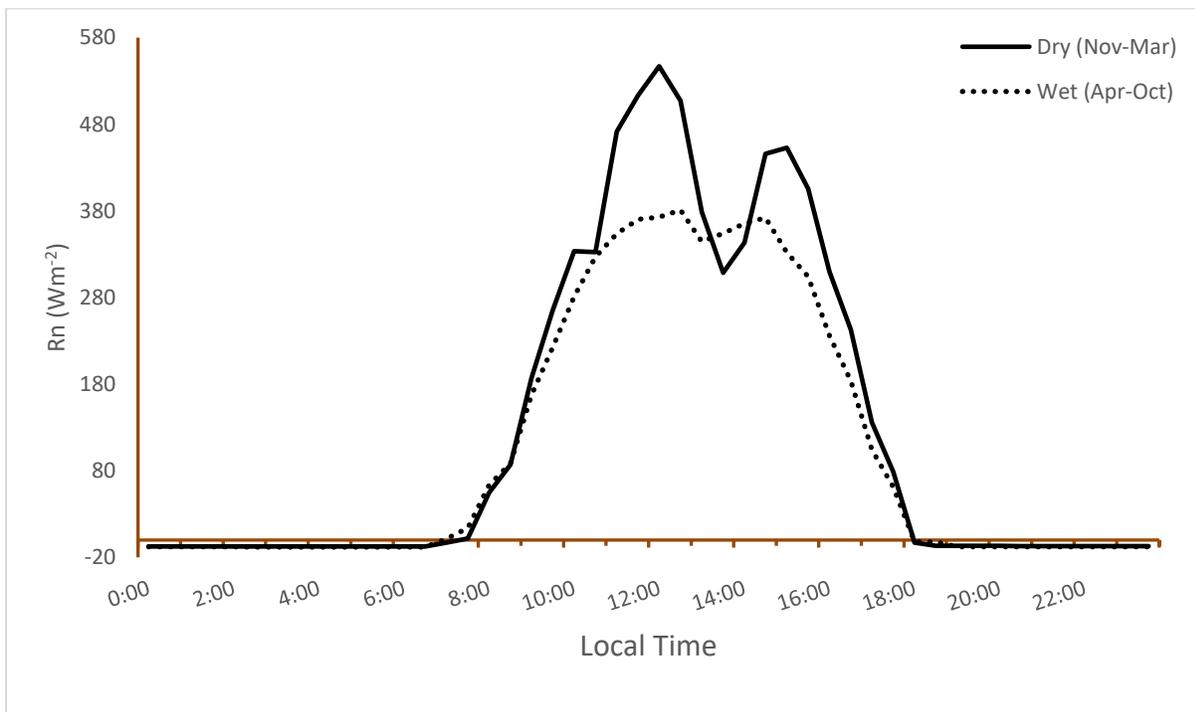

**Figure 1**: Average 30-minute diurnal variation of Net Radiation flux at Akure for both January (dry) and July (wet) seasons

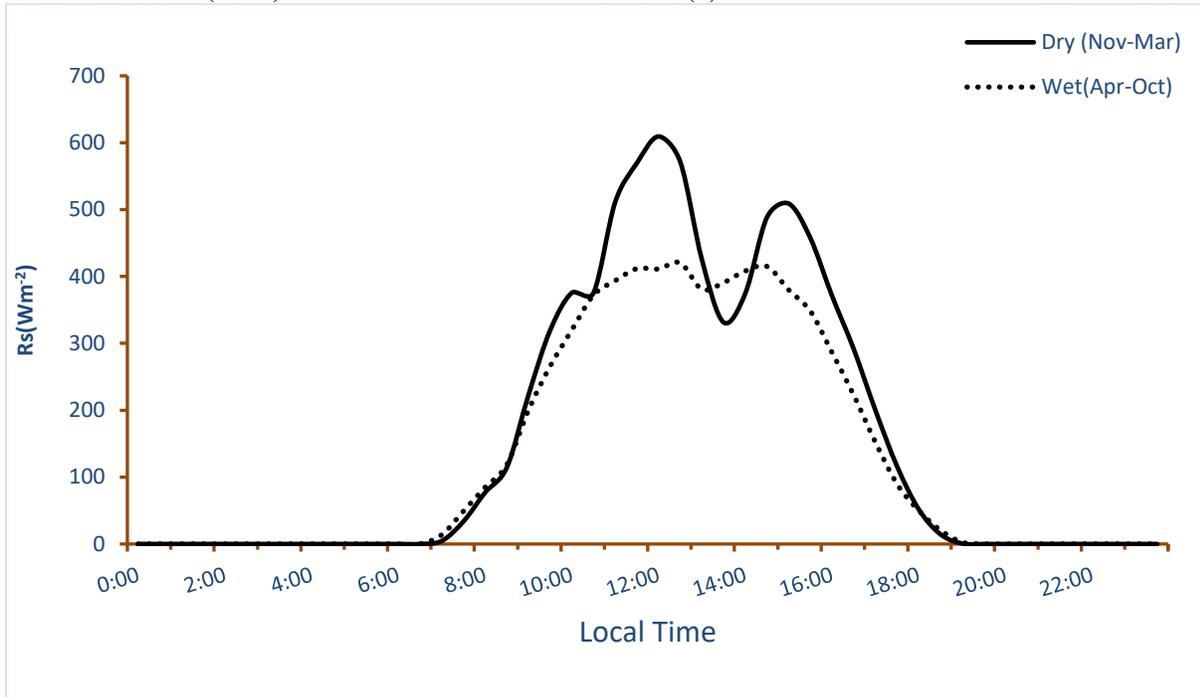

**Figure 2**: Average 30-minute diurnal variation of Solar Radiation flux at Akure in both January (dry) and July (wet) seasons

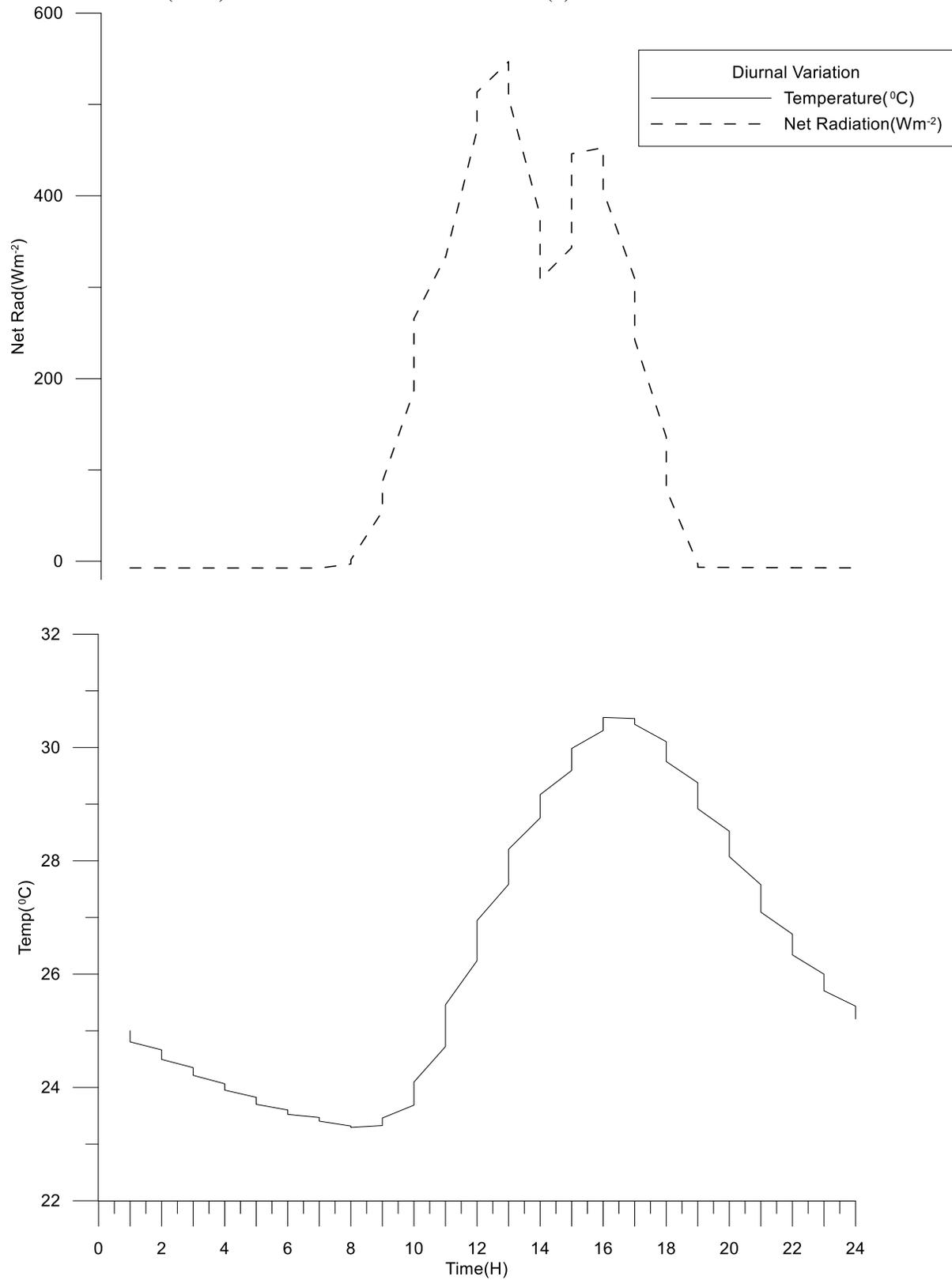

**Figure 3**: Average 30-minute diurnal Variation of Net Radiation flux and Temperature at Akure for the January dry season

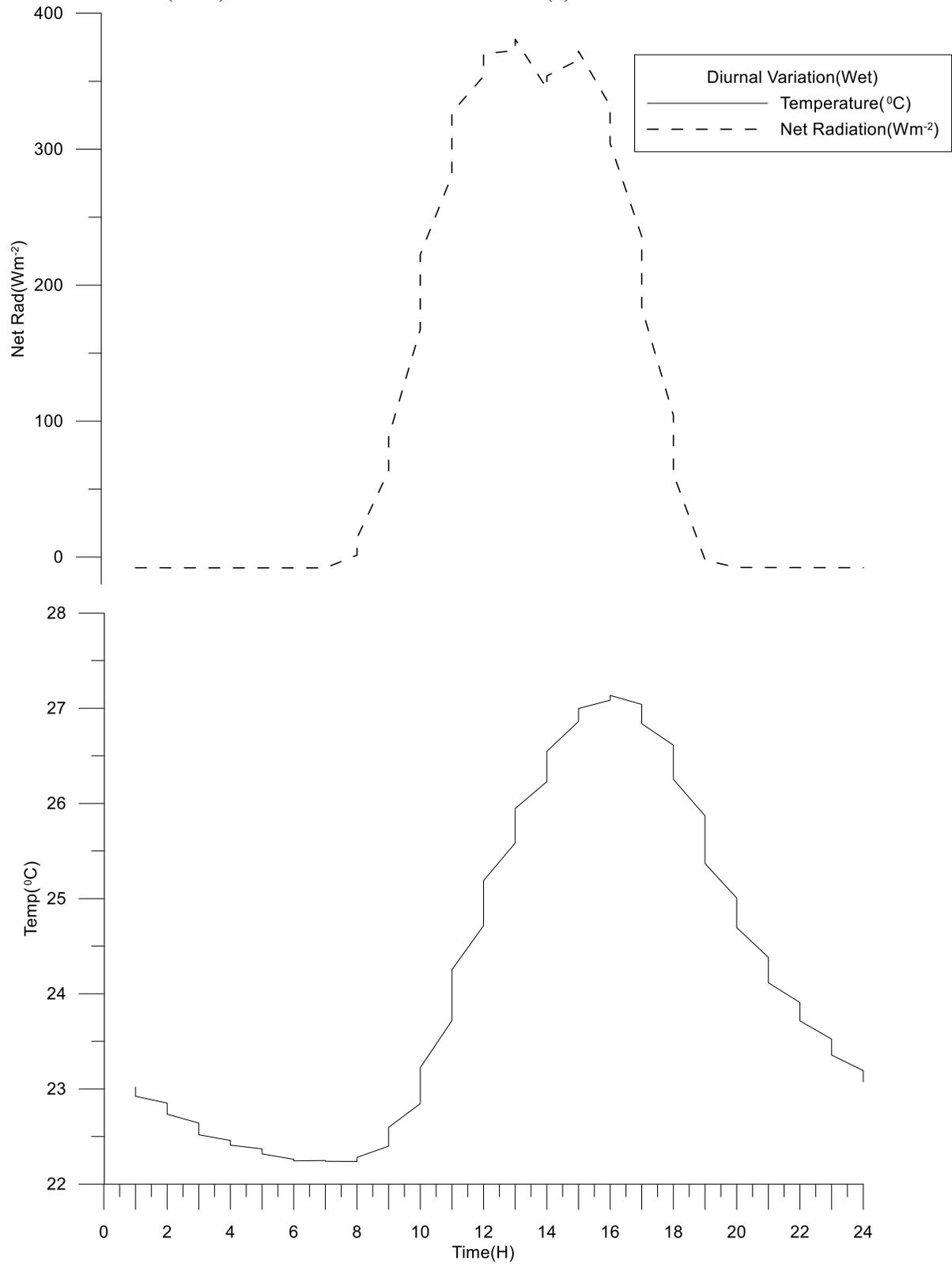

**Figure 4**: Average 30-minute diurnal Variation of Net Radiation flux and Temperature at Akure for the July wet season

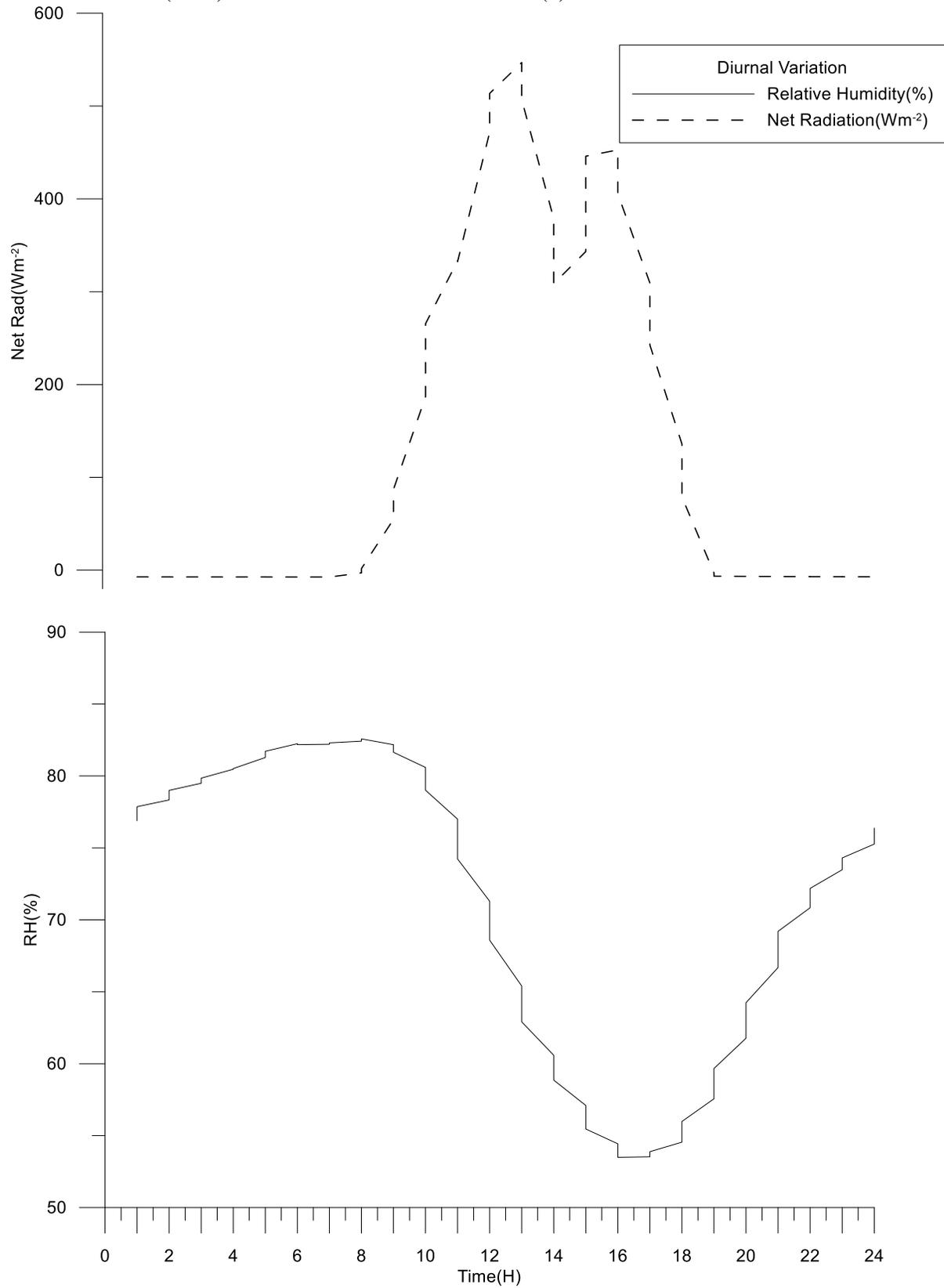

**Figure 5**: Average 30-minute diurnal variation of Net Radiation flux and Relative Humidity at Akure for the January dry season

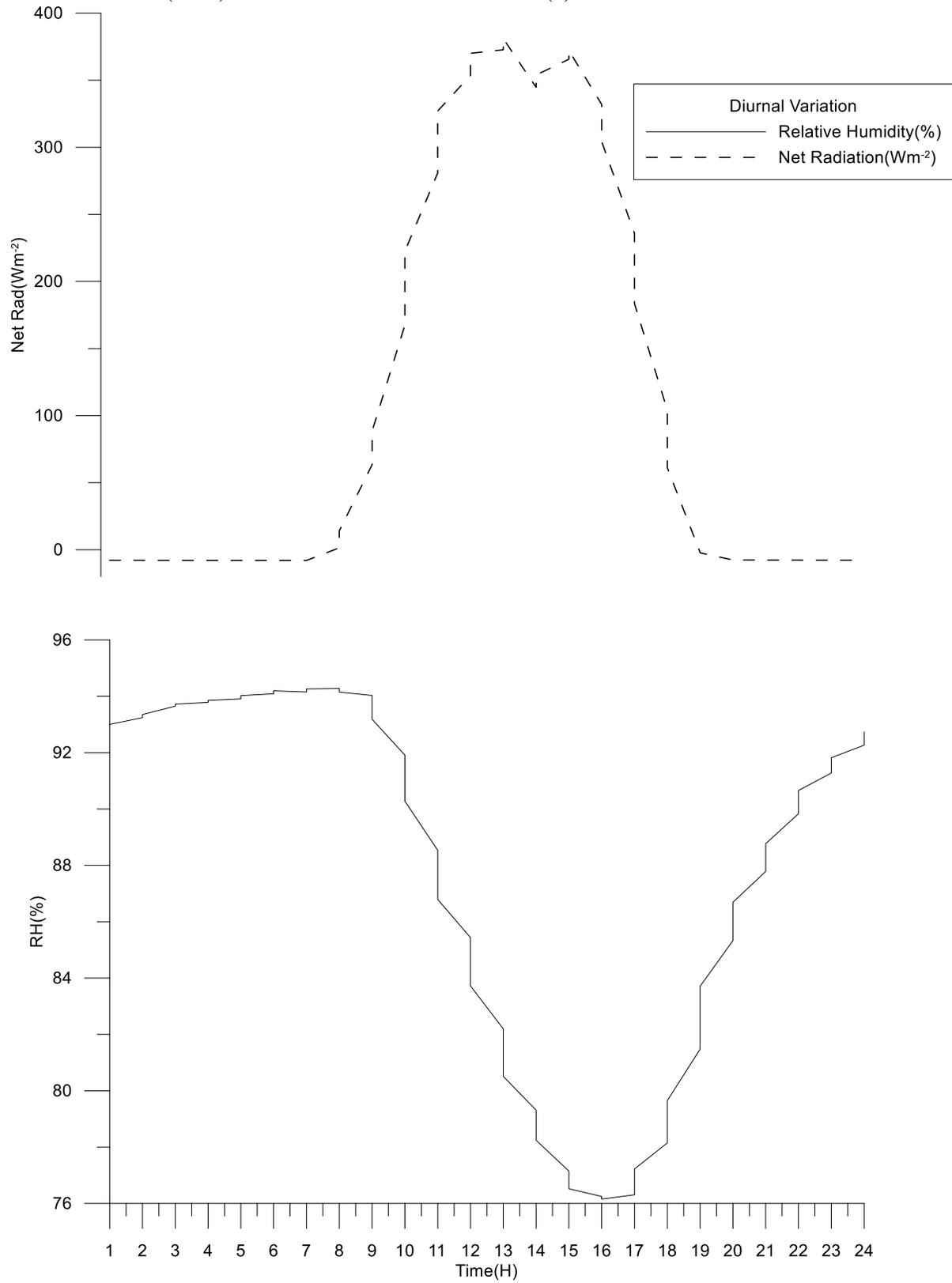

**Figure 6**: Average 30-minute diurnal variation of Net Radiation flux and Relative Humidity at Akure for July wet season

## 3.2 Daily Averages of Solar Radiation, Temperature, and Relative Humidity

The daily mean average of net radiation showed a bimodal distribution, as illustrated in Figure 7, with peak values of about 160 $Wm^{-2}$ $day^{-1}$ around February and November and a minimum of about 50 $Wm^{-2}$ $day^{-1}$ around July and August at the peak of the monsoon season. Large fluctuations in the daily averages were observed, especially during the wet months (April to October), as seen in Figure 7. This pattern can be attributed to the attenuating effects of clouds and aerosols on the incoming solar radiation, which vary vastly both on spatial and temporal scales, consequently affecting the radiation balance over the area [16].

The daily mean averaged solar radiation flux exhibited a similar bimodal pattern, as shown in Figure 8, with double peak values of about 200 $Wm^{-2}$ $day^{-1}$ around February and November and a minimum of about 60 $Wm^{-2}$ $day^{-1}$ around July and August.

Figure 9 depicts the daily average temperature, which showed a single peak, unlike net and solar radiation, occurring in February (the hottest month of the year at the measurement site), with the minimum in August. This difference in pattern highlights the complex relationship between temperature and radiation in the tropical climate.

Relative humidity (Figure 10) increased as the year progressed from January, reaching its peak around September-October. This trend can be primarily attributed to the climate of tropical West Africa, characterised by the interplay between monsoon and Harmattan winds, which meet at the Intertropical Convergence Zone.

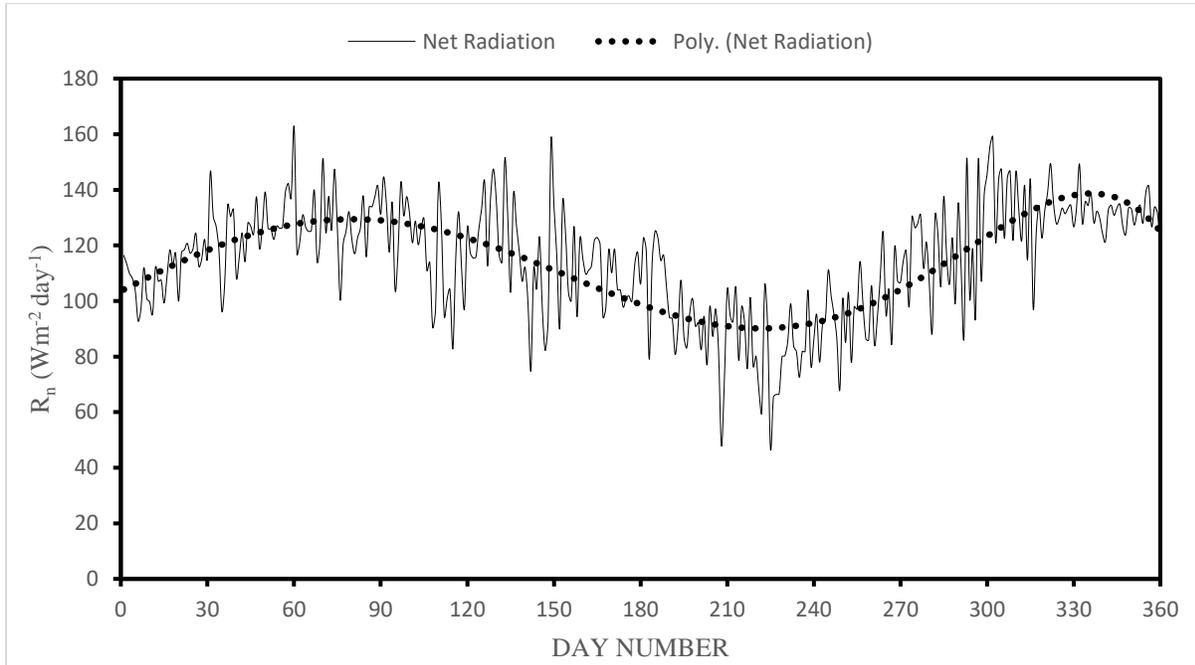

**Figure 7:** Daily Averages of Net radiation over Akure

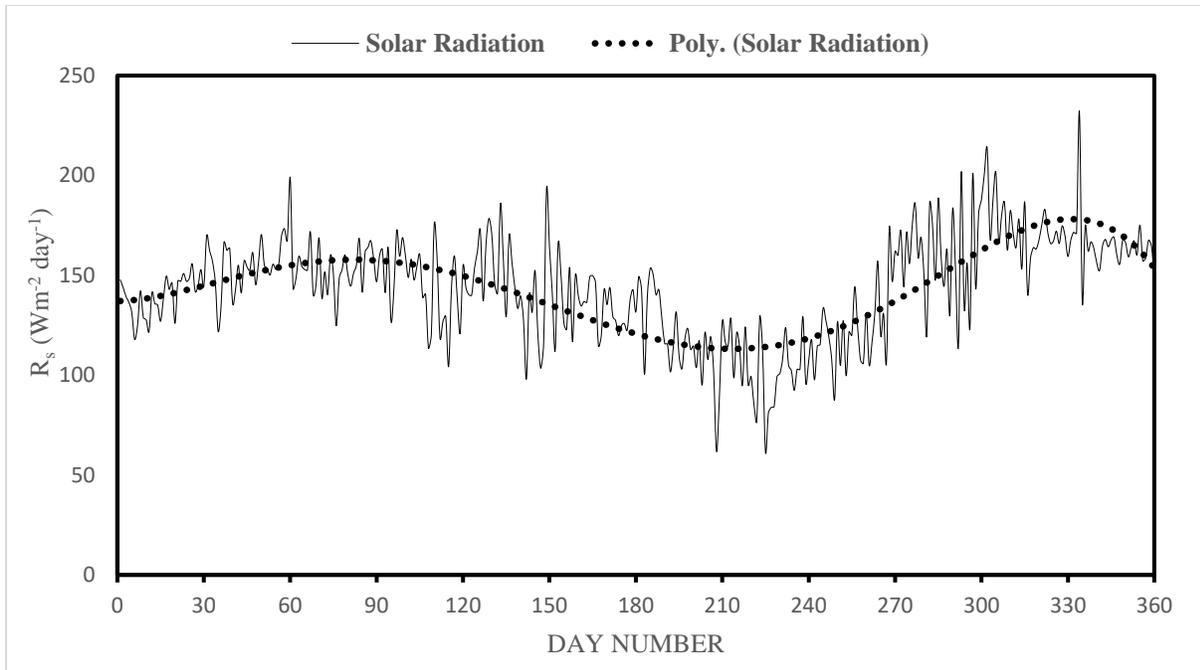

**Figure 8:** Daily averages of solar radiation over Akure

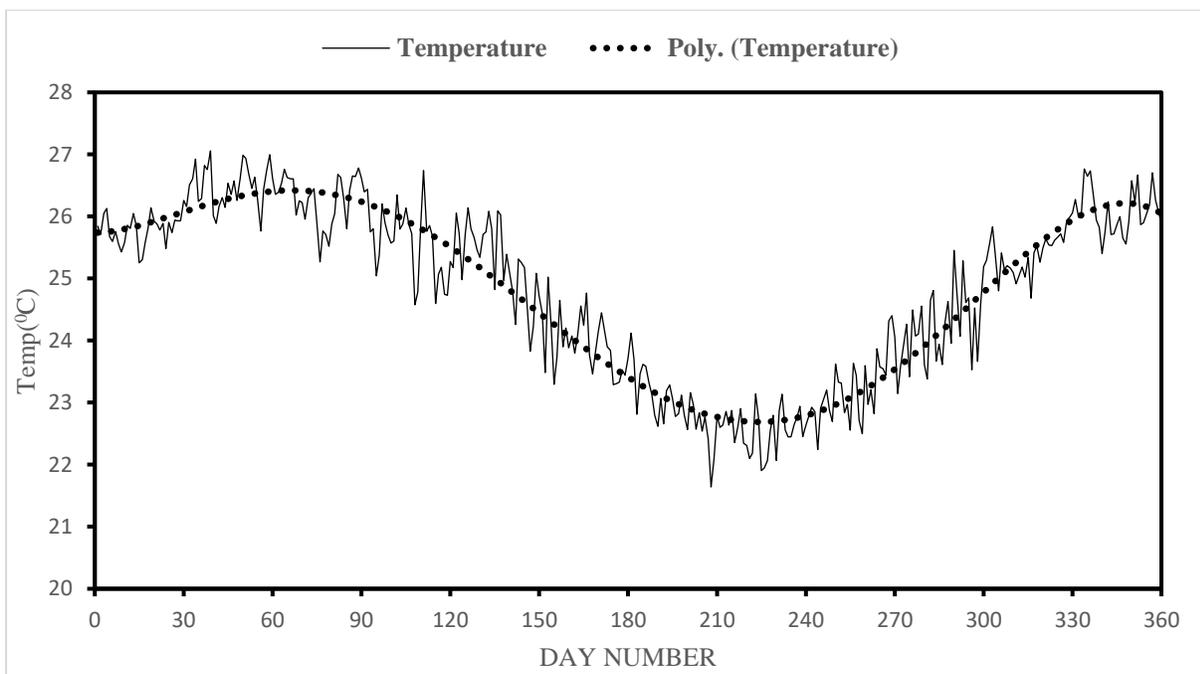

**Figure 9:** Daily average of Temperature over Akure

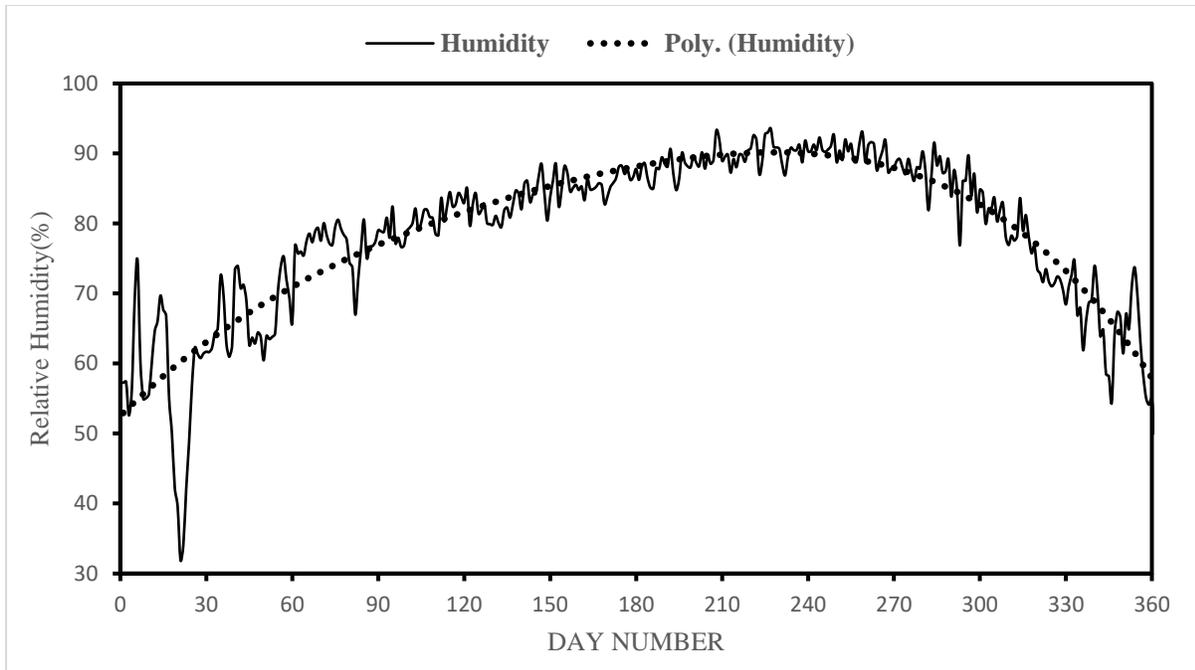

**Figure 10:** Daily Averages of Humidity over Akure

## 3.3 Monthly Averages of Net Radiation and Solar Radiation

Analysis of monthly averages revealed that the lowest mean solar radiation and net radiation occurred in August, while the highest values were recorded in November. As shown in Table 2, net radiation was a high percentage of solar radiation across all months, ranging from 74.1% to 84.0%. This high ratio indicates the significant role of solar radiation in determining net radiation in the tropical region.

The frequency distribution of daily averages showed that the higher percentage of the daily mean of net radiation was around 130 $Wm^{-2}$, while for solar radiation, it was around 150 $Wm^{-2}$. This indicates the high global solar radiation input experienced in the region, as reflected in the values presented in Table 2.

**Table 2:** Percentage of monthly Net radiation in Solar radiation for Akure (2007-2011)

| MONTHS | NET RAD ($Wm^{-2}$) | SOLAR RAD ($Wm^{-2}$) | PERCENTAGE (%) |
|---|---|---|---|
| JANUARY | 111.65 | 140.45 | 79.50 |
| FEBRUARY | 126.74 | 155.85 | 81.30 |
| MARCH | 128.30 | 152.70 | 84.00 |
| APRIL | 118.06 | 145.22 | 81.30 |

| MAY | 118.40 | 145.95 | 81.10 |
| JUNE | 111.54 | 136.34 | 81.80 |
| JULY | 94.39 | 116.75 | 80.80 |
| AUGUST | 82.03 | 102.51 | 80.00 |
| SEPTEMBER | 100.14 | 129.56 | 77.30 |
| OCTOBER | 123.29 | 166.43 | 74.10 |
| NOVEMBER | 132.28 | 170.28 | 77.70 |
| DECEMBER | 130.17 | 162.42 | 80.20 |

## 3.4 Comparisons of Estimated Daily $R_n$ from Different Models

We compared nine models with Jegede's [24] $R_n$ for Akure to evaluate their performance in estimating net radiation. As illustrated in Figure 11, most models had a fairly good correlation with Jegede's $R_n$ ($r^2 > 0.8$), with model 4 having the highest $r^2$ value ($r^2 = 0.92$) and model 9 with the lowest value ($r^2 = -1.02$).

Models 6, 7, and 8 compared poorly with Jegede's [24] $R_n$ calculations, with the lowest $r^2$ values, as evident from Figure 11. Models 6 and 7 had a cloudiness factor (f) of 1.0, which represents completely cloudy conditions. The effect of cloud cover errors can be subjective, although these conditions are known to attenuate $R_s$ reaching the earth's surface, which explains, in part, the underestimation of Jegede's [24] $R_n$ by these models.

Model 9, which used a multiple regression equation, performed poorly, as shown in Figure 11, likely due to the use of calibration coefficients suggested by Irmak et al. [2] that were only tested at mid-latitude sites. This underscores the importance of using locally calibrated coefficients for such models, as Holtslag and Van Ulden [1] suggested.

Model 1, which is the ASCE-EWRI [25] $R_n$ method, gave the best agreement with Jegede's $R_n$, recording the lowest divergence in Figure 11. This could be attributed to the fact that model 1 is a robust set of equations containing all the variables that affect net radiation.

All models performed better in the wet season than during the dry season when compared with Jegede's [24] $R_n$ a trend that can be observed in the scatter plots of Figure 11. This suggests that the models may be more accurate in higher humidity and cloud cover conditions, which are characteristic of the wet season in tropical regions.

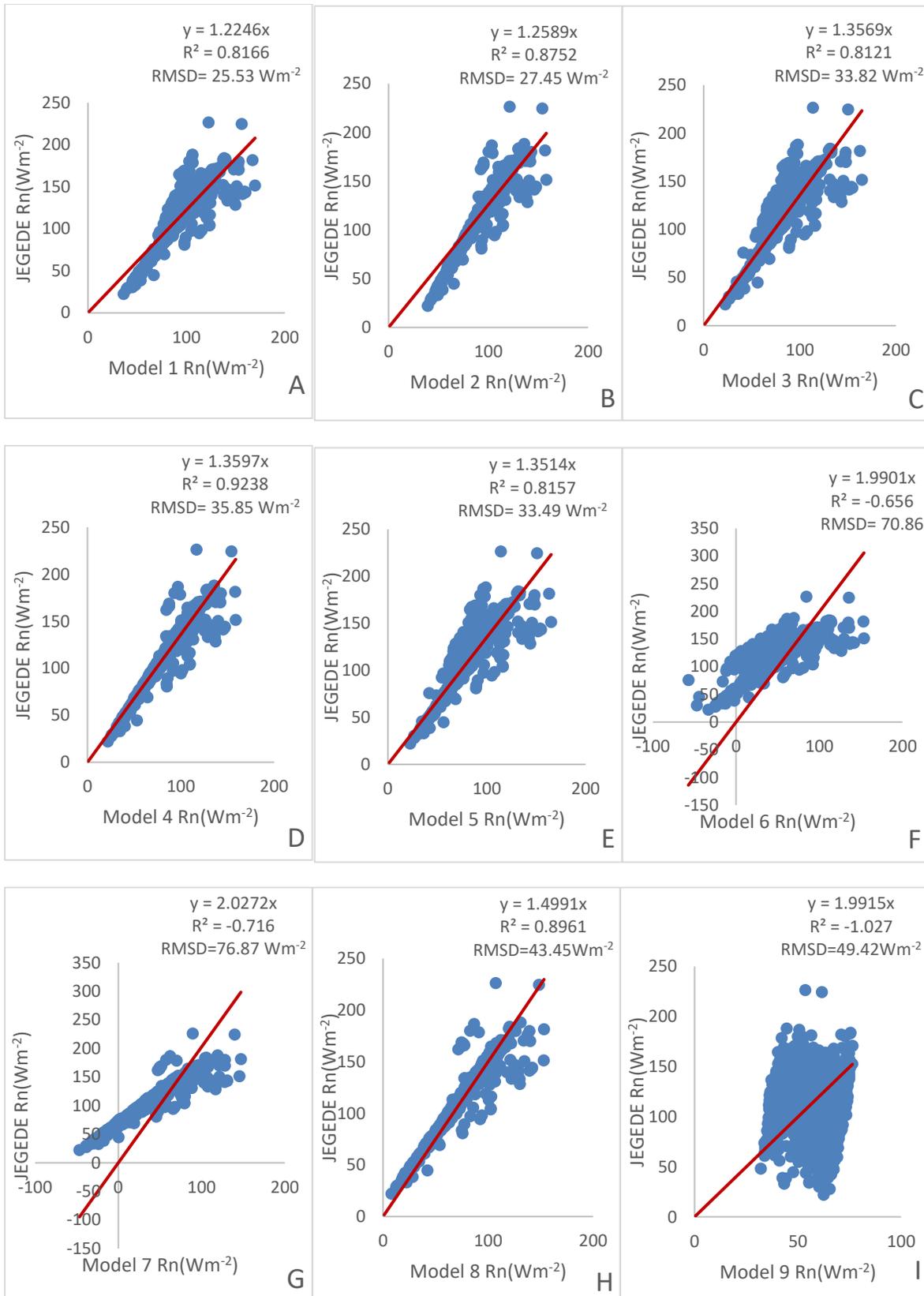

**Figure 11**: Relationship between the estimated daily Rn values using nine models and Jegede's estimated Rn values for Akure

### 3.5 Comparison Between Measured and Estimated Net Radiation

The standard model used for this work was compared with measured net radiation values. As shown in Figure 12, the diurnal variation of measured and estimated net radiation showed that although the estimated net radiation underestimates the measured net radiation, there is a good correlation between the two in the time window between 11:00 and 16:00 hours ($r^2 = 0.93$). This correlation is further illustrated in Figure 13, which presents a linear correlation between measured and estimated net radiation values.

These suggest the model performs well during peak radiation hours but may need refinement for early morning and late afternoon estimates.

These results, as visualised in Figures 12 and 13, highlight the complexity of estimating net radiation in tropical regions and the importance of considering local climatic conditions when applying or developing estimation models. The findings also underscore the need for continued in-situ measurements to improve and validate these models for specific geographic locations.

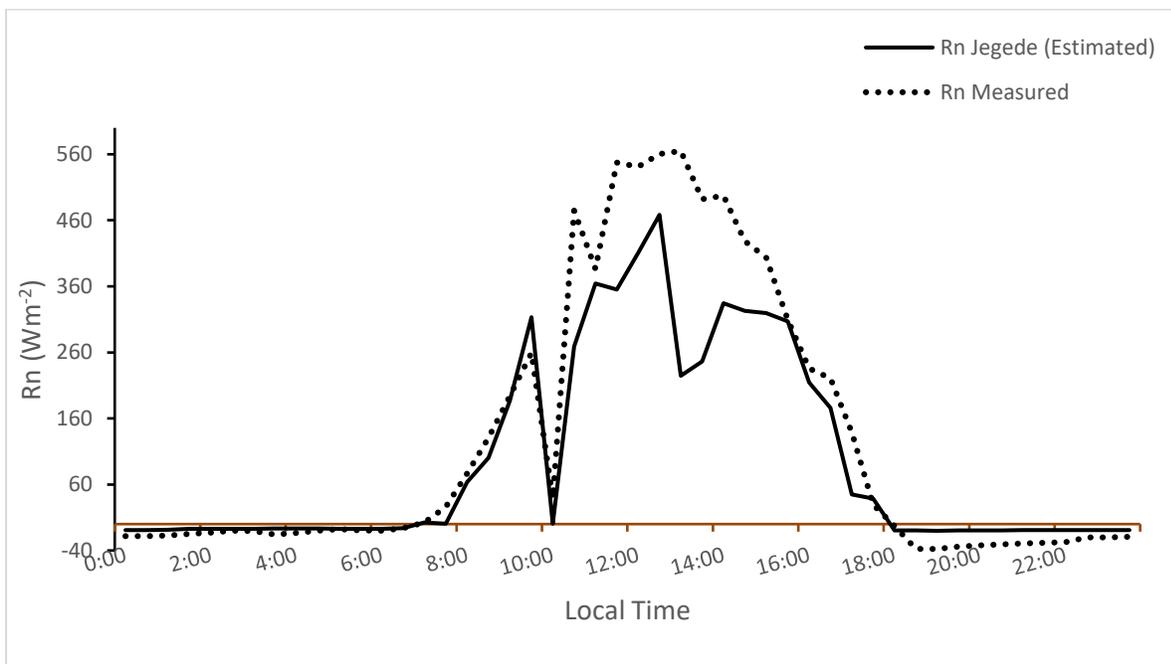

**Figure 12**: Comparison of measured net radiation with estimated net radiation

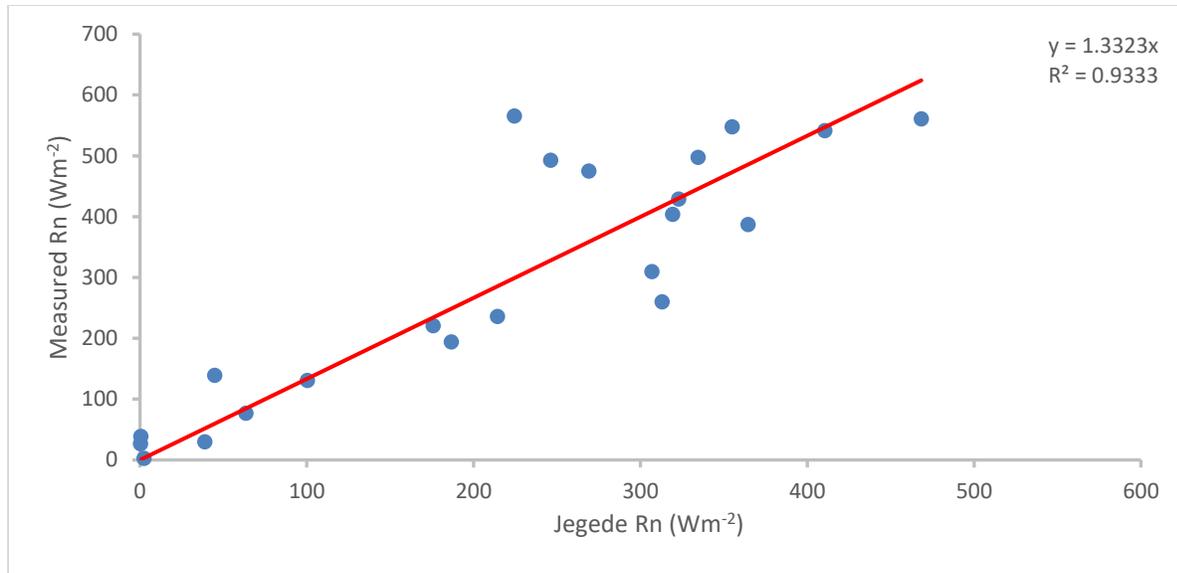

**Figure 13**: Linear correlation between measured and estimated net radiation values

## 4.0 Conclusion

This study has provided insights into the diurnal, monthly, and daily variation of net radiation in Akure, Nigeria. We found that net radiation exhibits a radiant energy surplus during the daytime and a deficit at night, with significant seasonal variations between wet and dry seasons.

The comparison of different models for estimating net radiation highlighted the importance of considering local conditions and cloud cover in these estimations. Models that were location-dependent or did not adequately account for cloud cover tended to underestimate net radiation compared to the standard model.

These findings contribute to understanding net radiation patterns in tropical West Africa and can inform future studies on the region's climate, agriculture, and water resource management.

**Declarations**

**Ethics approval and consent to participate**

Not applicable

**Consent for publication**

Not applicable

**Declaration of conflict of interests**

The authors declare no conflict of interest that could have influenced the work reported in this review.

**Funding**

No funding was sourced for this work.

**Authors' contributions**

All the authors drafted the manuscript read and approved the final manuscript.

**Acknowledgements**

The authors wish to thank the Communication/Atmospheric Physics Research Group of the Department of Physics at FUTA for permitting the use of their facility's data.